\documentclass[twocolumn]{aastex61}

\shorttitle{Oscillations under Coronal Holes}
\shortauthors{A.~Chelpanov, N.~Kobanov, M.~Chelpanov, A.~Kiselev}
\begin{document}
\title{Propagating Oscillations in the Lower Atmosphere under Coronal Holes}
\correspondingauthor{A.~\surname{Chelpanov}}
\email{chelpanov@iszf.irk.ru}
\author{Andrei~\surname{Chelpanov}}
\affil{Institute of Solar-Terrestrial Physics
                     of Siberian Branch of Russian Academy of Sciences, Irkutsk, Russia}
\author{Nikolai~\surname{Kobanov}}
\affiliation{Institute of Solar-Terrestrial Physics
                     of Siberian Branch of Russian Academy of Sciences, Irkutsk, Russia}
\author{Maksim~\surname{Chelpanov}}
\affil{Institute of Solar-Terrestrial Physics
                     of Siberian Branch of Russian Academy of Sciences, Irkutsk, Russia}
\author{Aleksandr~\surname{Kiselev}}
\affil{Institute of Solar-Terrestrial Physics
                     of Siberian Branch of Russian Academy of Sciences, Irkutsk, Russia}

\begin{abstract}

The subject of this study is oscillations in the lower atmosphere in coronal-hole regions, where the conditions are favorable for propagation between the atmospheric layers.
 Based on spectroscopic observations in photospheric and chromospheric lines, we analysed the features of the oscillations that show signs of propagation between layers of the solar atmosphere.
 Using the cross-spectrum wavelet algorithm, we found that both chromospheric and photospheric signals under coronal holes share a range of significant oscillations of periods around five minutes, while the signals outside of coronal holes show no mutual oscillations in the photosphere and chromosphere.
 The phase shift between the layers indicates a predominantly upward propagation with partial presence of standing waves.
 We also tested the assumption that torsional Alfv\'en waves propagating in the corona originate in the lower atmosphere.
 However, the observed line-width oscillations, although similar in period to the Alfv\'en waves observed earlier in the corona of open-field regions, seem to be associated with other MHD modes.
If we assume that the oscillations that we observed are related to Alfv\'en waves, then perhaps this is only through the mechanisms of the slow MHD wave transformation.

\end{abstract}

\section{Introduction} \label{sec:intro}

Coronal holes are areas of low plasma density and relatively low temperature in the outer atmosphere of the Sun.
They are associated with magnetic field rapidly expanding with height and the acceleration of the high-speed solar wind \citep{1975SoPh...40..351W, 2019ARA&A..57..157C}.

Coronal holes predominantly reside above unipolar areas.
The field extending from them forms the interplanetary magnetic field, and the outflowing plasma develops into the fast solar wind \citep{1973SoPh...29..505K,2002ESASP.508..361C,2009SSRv..144..383W}.
The acceleration of the fast solar wind occurs in the transition zone and the lower corona \citep{2005Sci...308..519T}.

The energy flux entering the chromosphere and the transition region, required to maintain the temperature of the corona and accelerate the fast solar wind, should be about 5$\times10^5$\,erg cm$^{-2}$\,s$^{-1}$ \citep[see reviews by][]{1978ARA&A..16..393V, 1981ARA&A..19....7K, 1990SSRv...54..377N,  1996SSRv...75..453N, 1991PPCF...33..539B,1993SoPh..148...43Z,2000PPCF...42..415J}.
Magnetohydrodynamic (MHD) waves dissipating in the upper solar atmosphere may be responsible for transporting a part of this energy.

In the matter of the energy transfer, one of the promising agents may be Alfv\'en waves.
Mainly open magnetic-field configuration in coronal holes provides favorable conditions for their propagation to considerable altitudes \citep{1998A&A...339..208B}.
Alfv\'en waves are thought to be caused by reconnections in the network; they contribute to the turbulence of the plasma flow from coronal holes \citep{1992sws..coll....1A,1993SoPh..144..155A,2018EGUGA..20.1790M}.
\citet{2007ApJS..171..520C} showed that Alfv\'en waves can be generated by granular motions \citep{2009SSRv..144..383W}.
Using \emph{Solar Dynamics Observatory}
(SDO) data, transverse oscillations were shown to be observed in coronal holes in the lower corona \citep{2011Natur.475..477M, 2014ApJ...790L...2T, 2018ApJ...852...57W}.
Nonthermal broadening of spectral lines has also been used as an indication of propagating torsional Alfv\'en waves in and under coronal holes \citep{1990ApJ...348L..77H, 2009A&A...501L..15B, 2012ApJ...751..110B, 2013ApJ...776...78H,2014AstL...40..222Z, 2016AstL...42...55K}.

Sources of the solar wind are located at the heights of the chromosphere and the transition zone \citep{2018EGUGA..20.1790M}, and yet most of the research on coronal holes concerns only their manifestations and characteristics in the upper atmosphere \citep[e.g.][]{2009A&A...499L..29B, 2020arXiv201208802B, 2009A&A...501L..15B, 2010ASSP...19..281B,2011SSRv..158..267B,2014ApJ...789..118K}, while the works studying the lower atmosphere under coronal holes have not been as numerous in the recent decade \citep{2003SoPh..217...53K, 2007Sci...318.1574D, 2007SoPh..243..143T, 2007ARep...51..773K, 2010SoPh..262...53T, 2014Sci...346A.315T, 2014AstL...40..222Z, 2016SoPh..291.1977G}.
With this article, we try to contribute to this area by analyzing the oscillations that we observe in the lower atmosphere under two coronal holes.

\section{Instruments and Data}

We carried out spectral observations for this work with the use of the ground-based \emph{Horizontal Solar Telescope} at the \emph{Sayan Solar Observatory} \citep{2004ARep...48..954K,2011SoPh..268..329K}.
The telescope allows for a spatial resolution of around 1$''$ and spectral resolution 4 to 15\,m{\AA}\ per pixel for the lines used in the observations; the temporal resolution of the series is four seconds.
The slit of the spectrograph covers a 25-arcsecond long region on the Sun’s surface.
During recording, a photoelectric guide moves the image to compensate for the Sun’s rotation.
From the spectrograms, we derived intensity, line-of-sight (LOS) velocity, and line-width signals.

The spectral lines that we used in the observations are the  Si\,\textsc{i} 10827\,{\AA}\ and H$\alpha$ lines that form in the photosphere and chromosphere, respectively.

To analyze the oscillation spectral composition, we used the fast Fourier transform algorithm and the Morlet wavelet.
We calculated the confidence levels using the technique described by \citet{1998BAMS...79...61T}.
It is based on comparing dynamic spectrum characteristics with a theoretical background noise.
The method implies a white- or red-noise spectrum, although more complicated models are also used \citep{2016ApJ...825..110A}. The white-noise model is implied for our data, as the wavelet power stays roughly constant over the range of periods of interest 2\,--\,10\,minutes.

For additional analysis, we used data from the \emph{Atmospheric Imaging Assembly} \citep[AIA:][]{sdoaia} onboard the \emph{Solar Dynamics Observatory} (SDO). The instrument provides full-disk images of the Sun in several ultraviolet channels with a 0.6$''$ spatial resolution and a temporal cadence as short as 12\,seconds.
To compensate for the rotation, we used the algorithm available in the SunPy core package \citep{sunpy_community2020}.

\section{Results and Discussion}

\subsection{Intensity and LOS Velocity Oscillations}

\begin{figure}
\centerline{
\includegraphics[width=7cm]{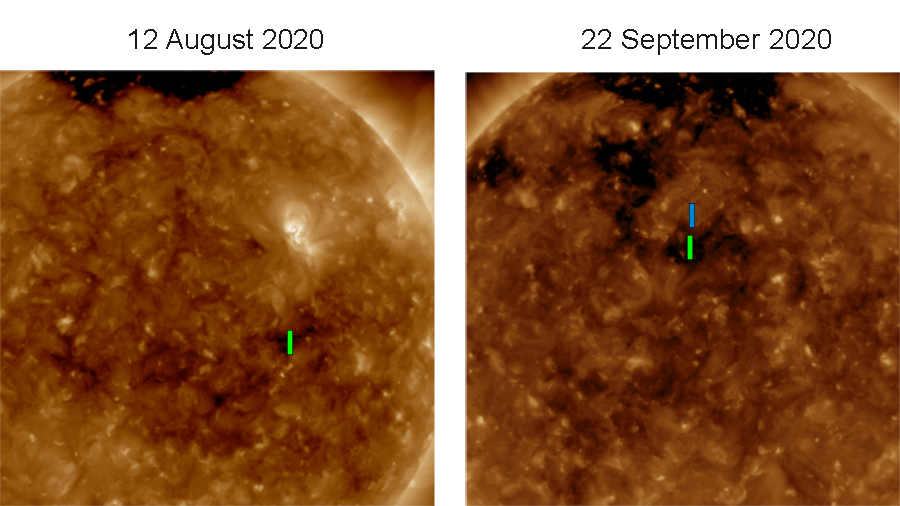}
}
\caption{The locations of the spectrograph slit in the coronal-hole regions superimposed on the 193\,{\AA} channel images. The second panel also shows the location of the slit in a region outside the coronal hole for the quiet-Sun series.}
\label{fig:CH-193}
\end{figure}

We carried out spectral observations in the  Si\,\textsc{i} 10827\,{\AA}\ and H$\alpha$ lines in the bases of two coronal holes on 12 August 2020 (80-minute series) and 22 September 2020 (105\,minutes). Figure~\ref{fig:CH-193} shows the locations of the coronal holes.
These lines form in the upper photosphere and chromosphere respectively.
Immediately after recording the coronal hole series on 22 September, we recorded a similar series outside of the coronal-hole region close to it to compare the oscillation parameters in the coronal hole and in a quiet-Sun region.

\begin{figure}
\centerline{
\includegraphics[width=8cm]{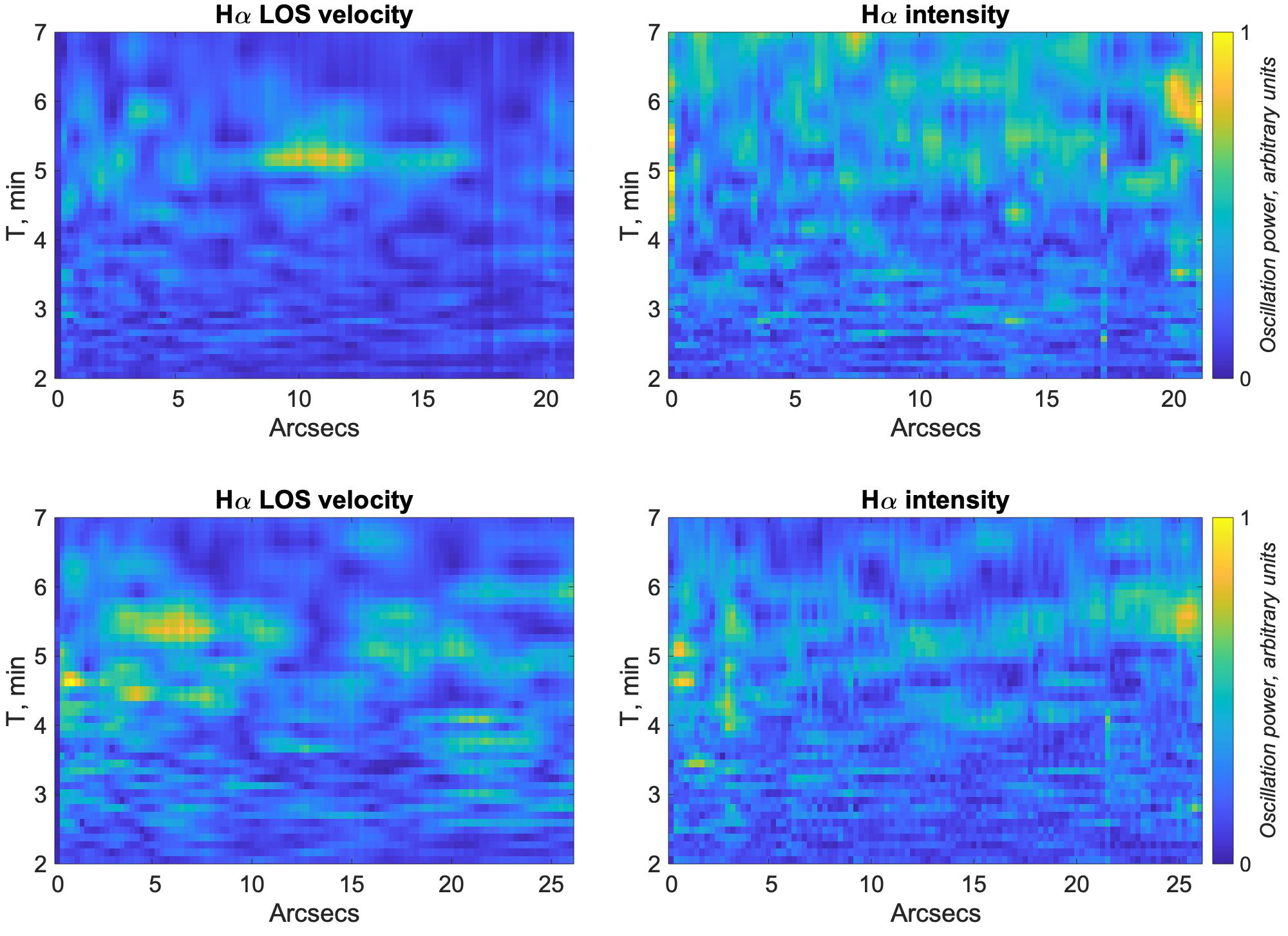}
}
\caption{Oscillation power distribution along the slit in the first (\emph{upper row}) and second (\emph{bottom row}) coronal holes.}
\label{fig:along-the-slit}
\end{figure}

The distribution of the Fourier oscillatory power over the length of the slit in the coronal holes (Figure~\ref{fig:along-the-slit}) shows that the power is evenly distributed along the slit with areas of a slight increase. The maximum power is at 10$''$ for the first coronal hole and at 6$''$ for the second.

For the analysis, we used wavelet spectra, which show frequencies dominating in the signal over the recording time of the series.
We constructed cross-spectra based on pairs of signals to show similar frequencies at the same temporal intervals in these signals.
They are calculated by multiplying the wavelet spectrum of one series with the complex conjugate of another.
Cross-spectra indicate the intersections of the time--frequency domains that show oscillation power in two series.

\begin{figure}
\centerline{
\includegraphics[width=8cm]{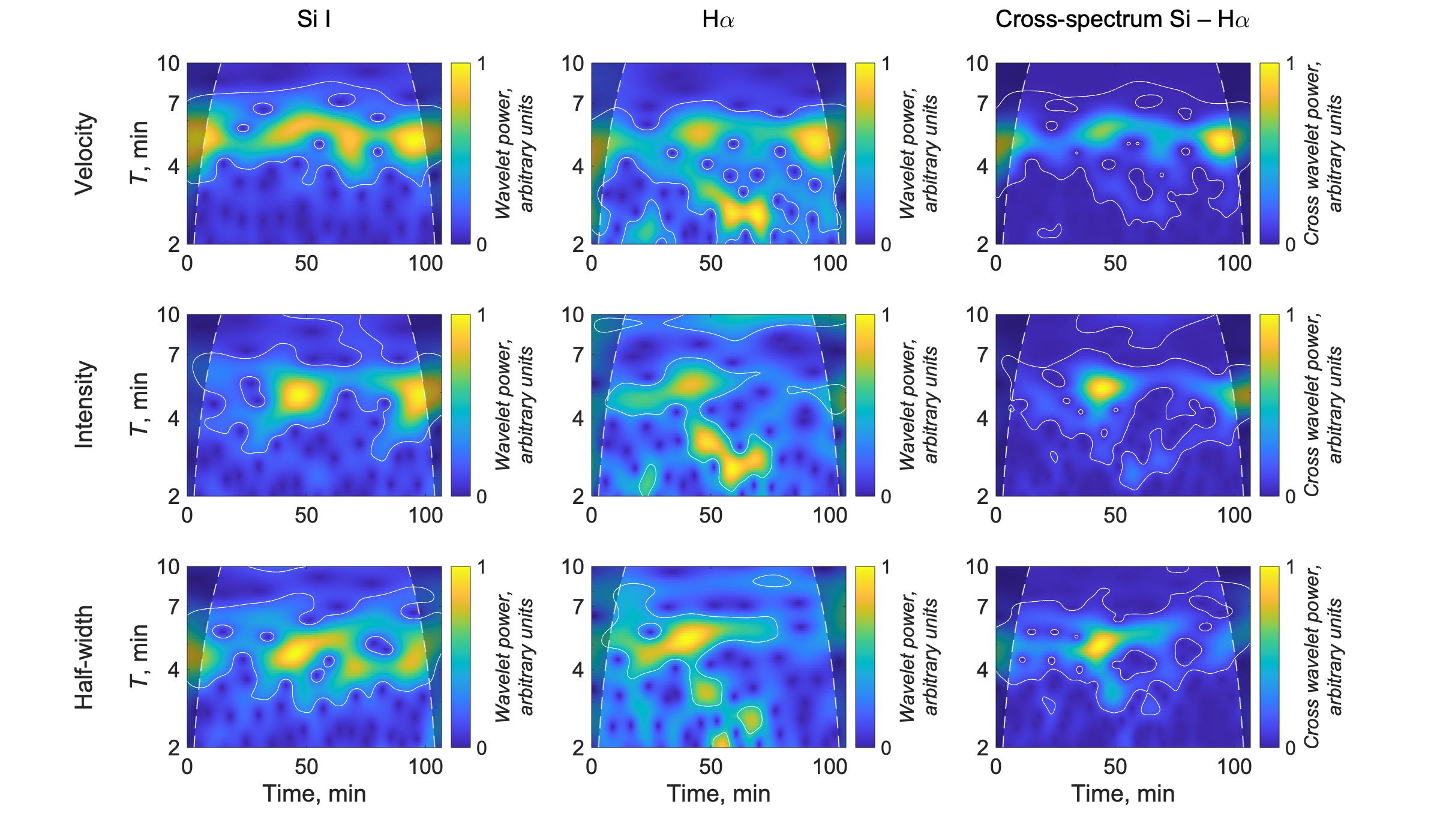}
}
\caption{Wavelet spectra and cross-spectra in the coronal hole on 22 September 2020 for different parameters in the H$\alpha$ and Si\,\textsc{i} lines in the six arcsecond point. The regions enclosed by the light contours ans dashed lines have significance greater than 99\,\% relative to the corresponding white noise.}
\label{fig:CH-wvl}
\end{figure}

We used cross-spectra to determine frequencies that exist simultaneously at two altitude levels -- in the photosphere and chromosphere (Figure~\ref{fig:CH-wvl}), and therefore can propagate upwards (or downwards) in the atmosphere of the coronal-hole region.

\begin{figure}
\centerline{
\includegraphics[width=8cm]{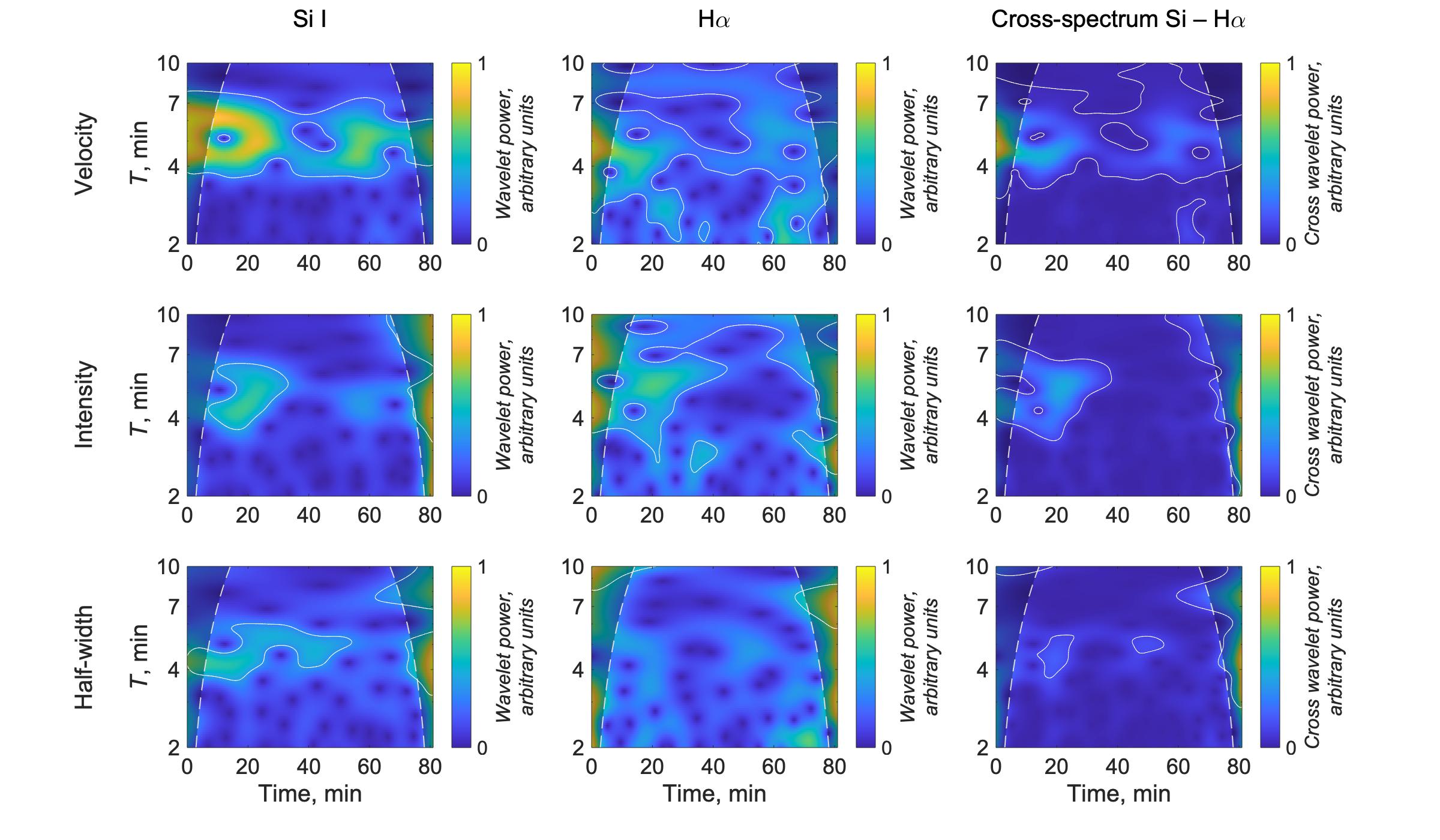}
}
\caption{Wavelet spectra and Si\,\textsc{i}--H$\alpha$ cross-spectra for the quiet-Sun series on 22 September 2020 for different parameters in the H$\alpha$ and Si\,\textsc{i} lines. The regions enclosed by the light contours ans dashed lines have significance greater than 99\,\% relative to the corresponding white noise.}
\label{fig:QS-wvl}
\end{figure}

For comparison, Figure~\ref{fig:QS-wvl} shows similar spectra typical of the quiet-Sun region series.

Oscillations with periods of four to five minutes dominate both in the photosphere and in the chromosphere of the coronal holes in all of the observed parameters.
This could indicate that these are the periods of waves propagating between the studied atmosphere layers.
By comparison, in the region of the quiet Sun, oscillations at different heights are observed at different frequencies, and the cross-spectra do not show mutual high-power domains.
In this case, one can say that direct propagation of oscillations is not observed, in contrast to coronal holes.

We determined the periods of maximum power on the diagrams of mutual-period oscillations in the coronal-hole regions, averaged over the observation area.
The maximum power of the cross-spectra falls in the ranges of periods 1.6\,minutes wide centered at 4.83 and 5.06\,minutes for the first and second coronal holes.
Earlier, \citet{2010SoPh..262...53T} also noted the prevalence of five- over three-minute oscillations in the Ca\,\textsc{ii} K and 8498\,{\AA}\ lines under coronal holes.

\begin{figure}
\centerline{
\includegraphics[width=8cm]{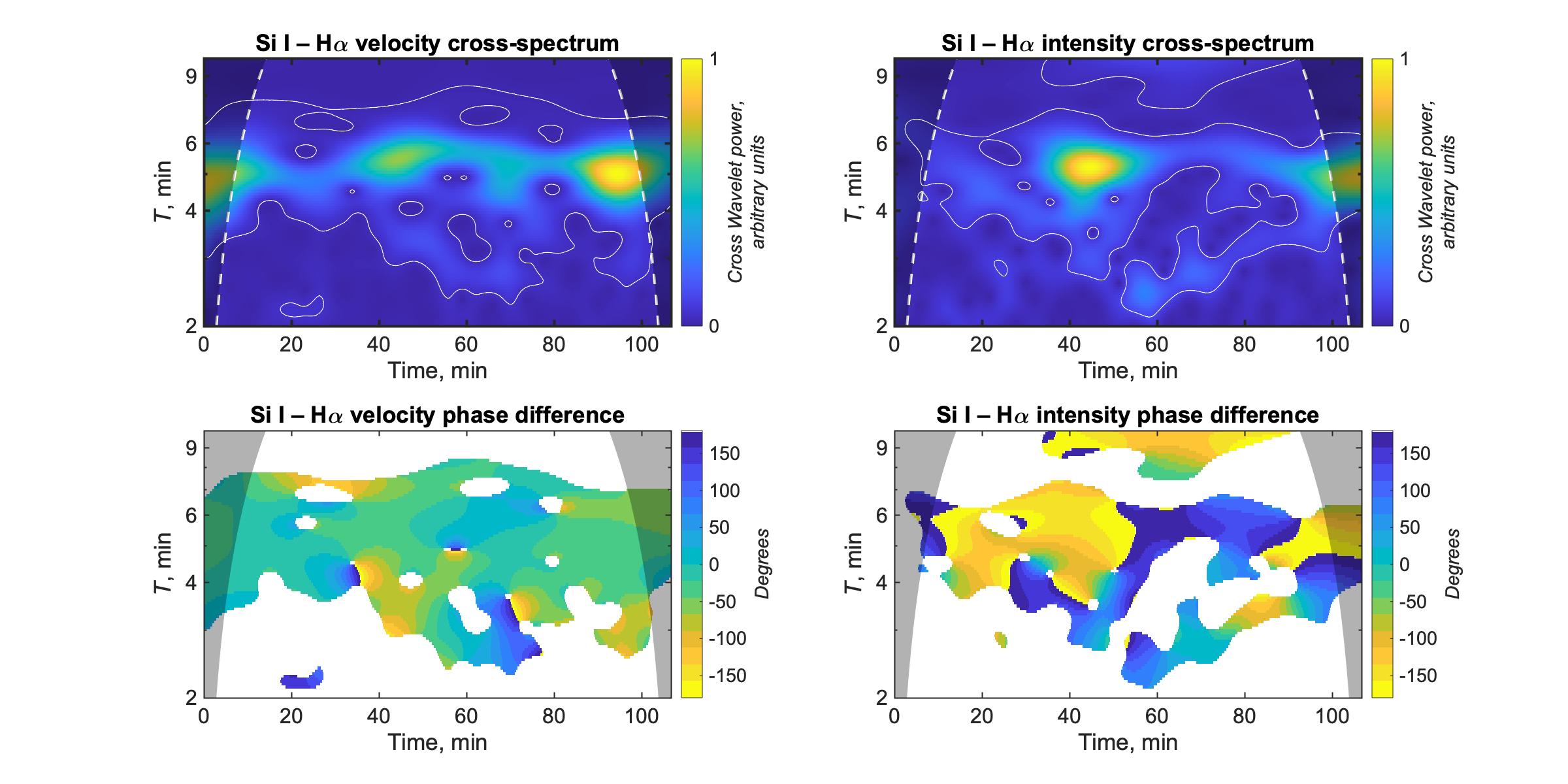}
}
\caption{\textit{Upper row}: cross wavelet power between the H$\alpha$ and Si\,\textsc{i} velocity (left) and intensity (right) signals. \textit{Bottom row}: phase differences between the signals. The areas outside of the 99\,\% significance contour are filled with white and/or shaded.}
\label{fig:V-I-phase}
\end{figure}

To estimate the parameters of wave propagation between the layers of the solar atmosphere, we measured the phase difference between the signals.
Figure~\ref{fig:V-I-phase} shows the values of the phase difference between the Si\,\textsc{i} and H$\alpha$ intensity signals as well as the Si\,\textsc{i} and H$\alpha$ velocity signals throughout the series for different periods.

We estimated the average value of the phase delay for significant oscillations over all spatial points of the slit.
For the velocity signals, the value of the lag between the photosphere and chromosphere in the 5\,$\pm$\,0.5\,minute period range is 22.6\,$\pm$\,12.8\,degrees.
The intensity signals show a more significant phase difference scatter.
This may indicate that the contribution of standing waves to the observed five-minute oscillations varies during the time series.
Another explanation may be absorption by unresolved chromospheric plasma nonuniformities \citep{2009ApJ...705..272R}.

The observed phase difference may help us to assess the propagation speed.
We assume the height difference between the formation levels of the two lines for the Si\,\textsc{i} and H$\alpha$ lines to be approximately 1\,Mm \citep{2008ApJ...682.1376B,2012ApJ...749..136L}. This gives an average speed of 54\,km\,s$^{-1}$. This speed is greater than the sound speed in the chromosphere; it, however, falls within the reasonable error range given the uncertainties in the difference between the formation heights and the measured time lag.

\begin{figure}
\centerline{
\includegraphics[width=8cm]{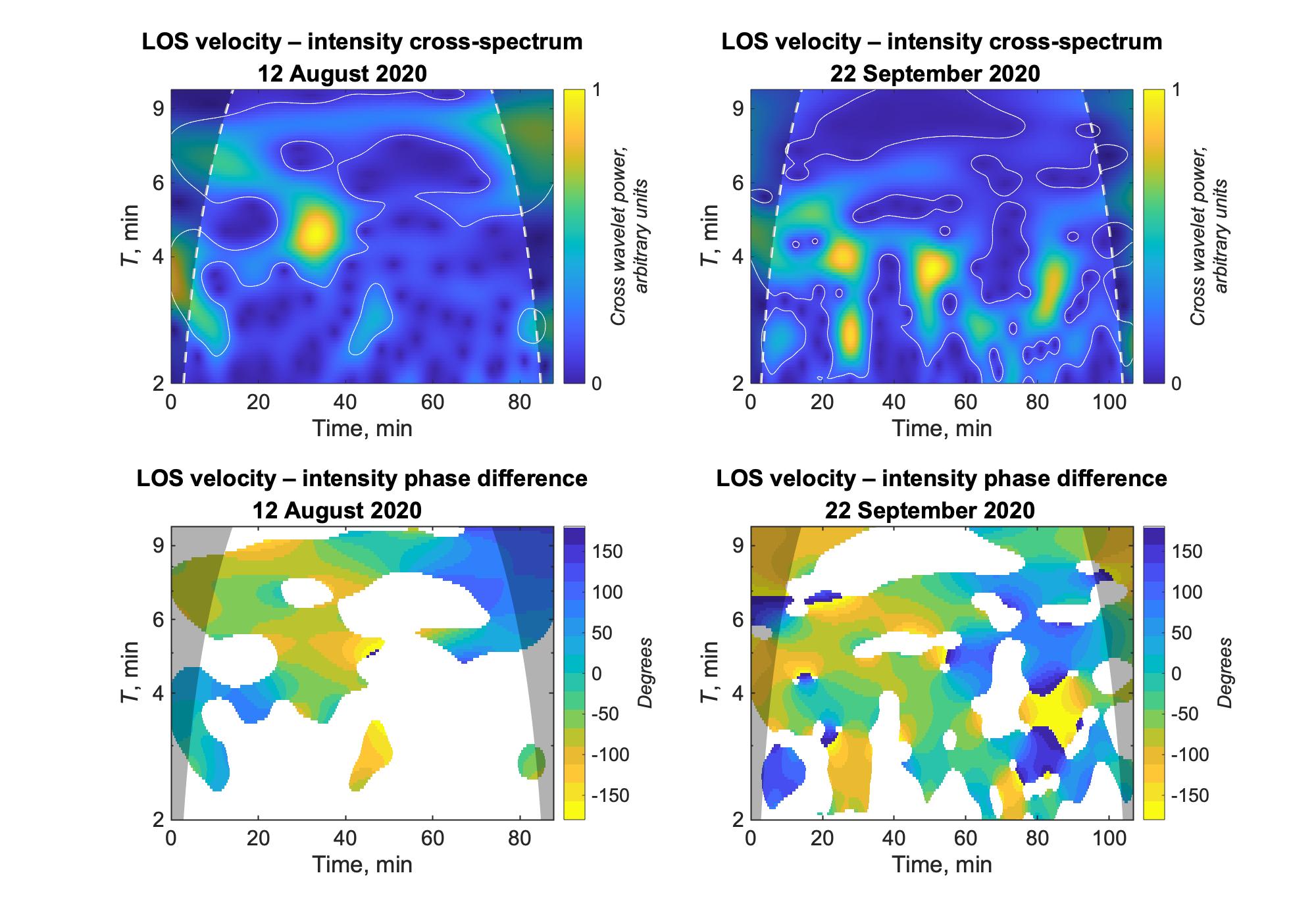}
}
\caption{Examples of the phase differences between the H$\alpha$ velocity and intensity signals in coronal-hole regions. The areas outside of the 99\,\% significance contour are filled with white and/or shaded.}
\label{fig:V-I-phase-2}
\end{figure}

To analyze the type of observed waves, one can compare the phase difference between the velocity and intensity signals of the same layer.
A zero phase shift between them indicates a propagating slow magnetoacoustic wave.
The diagrams in Figure~\ref{fig:V-I-phase-2} show that in the time-period domains of significant oscillations, a shift from 0 to about 100 degrees is observed.
This may confirm the assumption that the type of waves that exist at the observed heights changes during the time series.

Another type of MHD waves is the sausage mode.
When it is observed, the line-width signals should demonstrate a double frequency compared to the frequency observed in the intensity signal \citep{2013A&A...555A..74A,2017AstL...43..844K}.
However, this is not found in observations.
On the contrary, the frequencies of significant oscillations of these parameters coincide.

\begin{figure}
\centerline{
\includegraphics[width=8cm]{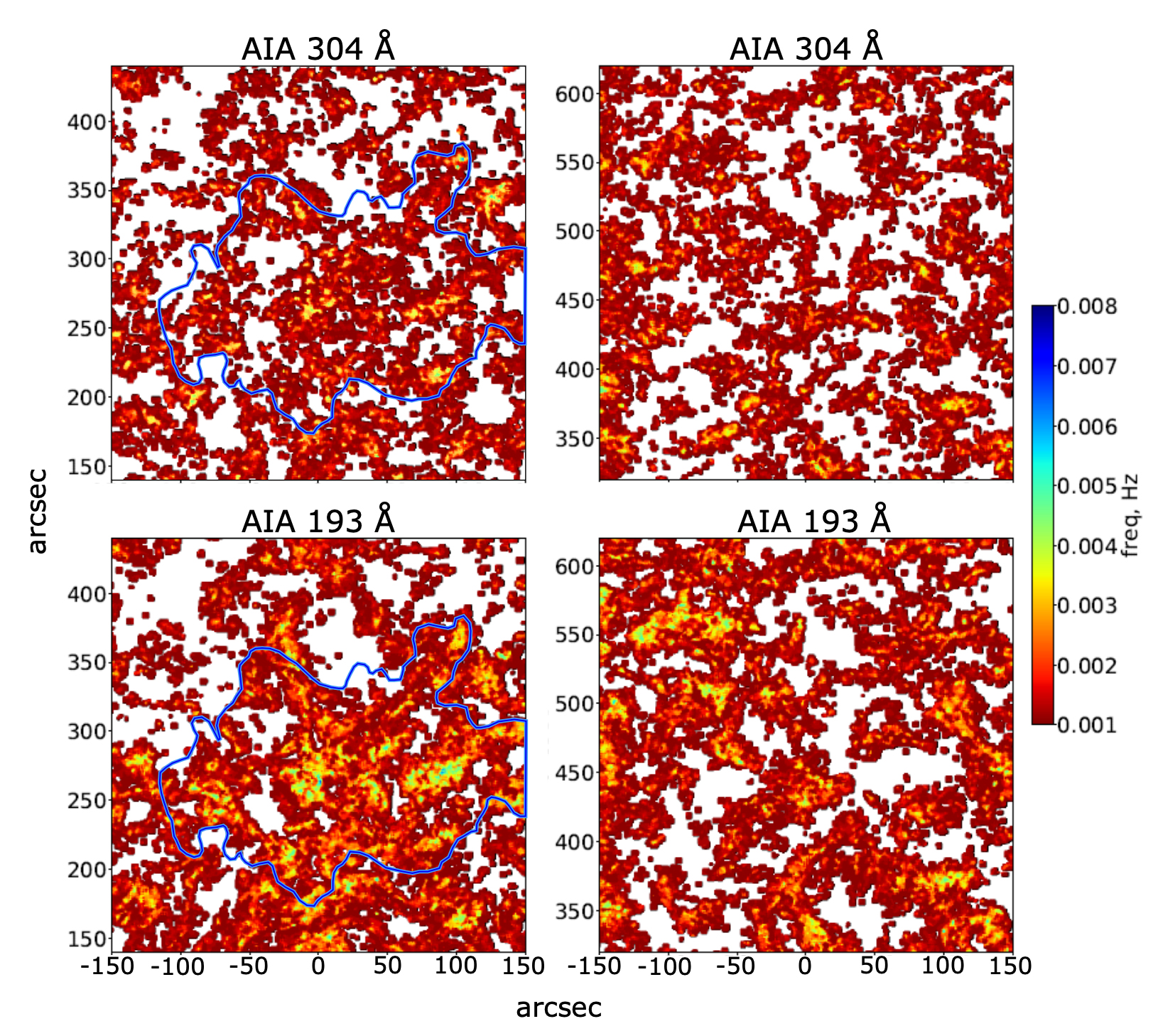}
}
\caption{\textit{Left}: distribution of dominant frequencies in the AIA channels in the coronal-hole region; the blue contour shows borders of the coronal hole as they appear in the 193\,{\AA}\ channel. \textit{Right}: distribution of dominant frequencies in the quiet-Sun region for comparison.}
\label{fig:SDO}
\end{figure}

The distribution of dominant frequencies in the SDO channels also shows a slight increase in oscillation power in the five-minute (or 3.3\,mHz) range in the region of the coronal hole with respect to the quiet Sun (Figure~\ref{fig:SDO}).
The dominant frequencies were derived from the FFT-spectra.
To clear the images of the noise-dominated areas, we applied an image-morphology method \citep{Serra1988image} based on the assumption that the values vary to a greater extent in the signal-dominated spatial points: a $5\times5$ window is passed across an image, and the standard deviation within the window is calculated.
Then, for the points where the standard deviation is close to zero, the central point is neglected and filled with white. After this procedure, only the points of the highest variation in the signal are left in the maps.
The resulting images were smoothed using morphological dilation with a $5\times5$ disk as the structuring element.

To a greater extent, this is seen at coronal heights (the 193\,{\AA}\ channel), but even at the height of the transition region, a denser concentration of significant oscillation areas is seen in the central part of the coronal-hole region.

\subsection{Line-Width Signals}

The observational manifestations of torsional Alfv\'en waves are usually associated with the line width oscillations available in our observations.
In cases where a magnetic tube is located at an angle to the line of sight, the rotating of the tube contributes simultaneously to the red and violet shifts of the spectral line, which leads to its broadening, while such oscillations do not affect the intensity and line-of-sight velocity signals.
Therefore, when observing true torsional Alfv\'en waves, synchronous radial-velocity or intensity signals should not accompany the line-width oscillations.

\begin{figure}
\centerline{
\includegraphics[width=8cm]{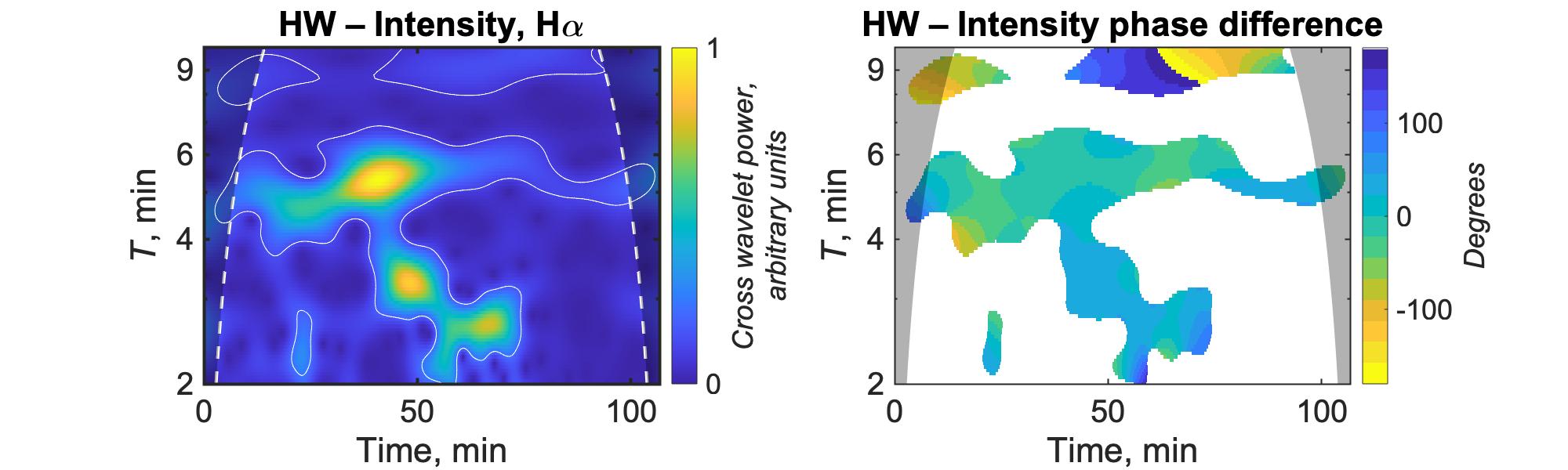}
}
\caption{\textit{Left}: cross-spectrum of the H$\alpha$ line-width and intensity signals in the coronal hole on 22 September 2020; \textit{right}: phase difference between the line-width and intensity signals is close to zero during the time series.}
\label{fig:HW-int-phase}
\end{figure}

In our observations, the periods of significant oscillations in the profile line-width signals are distributed in the range from four to six minutes (Figure~\ref{fig:CH-wvl}).
In the coronal holes that we observed, however, the power distribution in the time-period diagrams approximately coincides in the intensity and line-width signals.
In addition, the phase difference between oscillations in intensity and line width in the significant oscillations domains of the diagram remains close to zero during the observation time (Figure~\ref{fig:HW-int-phase}).

\begin{figure}
\centerline{
\includegraphics[width=8cm]{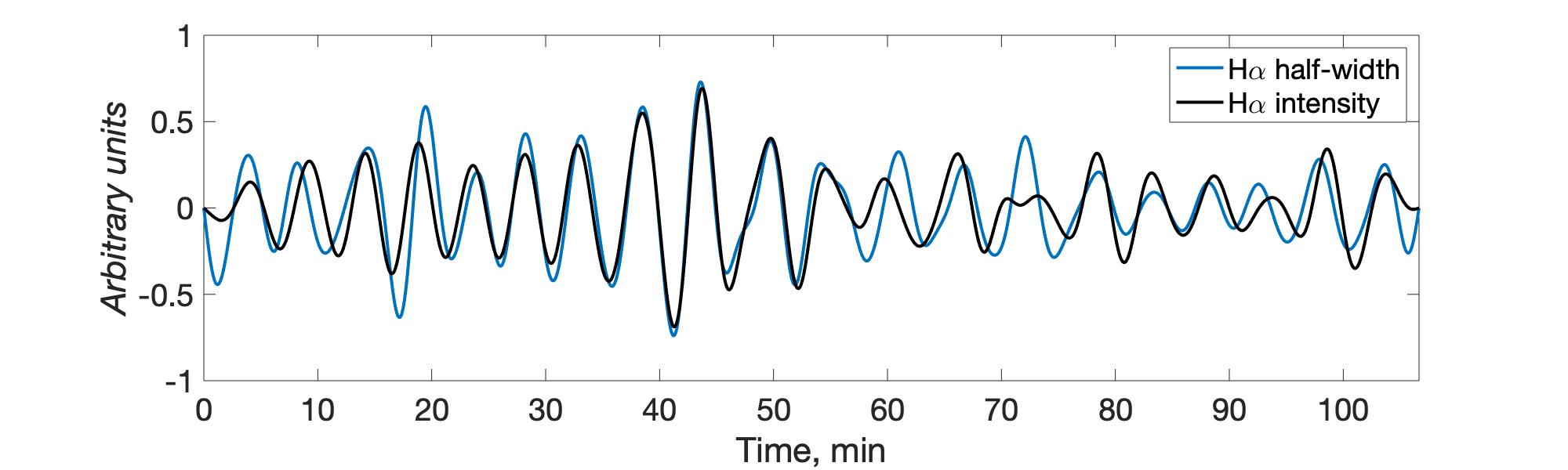}
}
\caption{Intensity and line-width signals of the of the H$\alpha$ line filtered in the 4.5\,--\,5.5 minute range in a coronal-hole region.}
\label{fig:wave-trains}
\end{figure}

In the wave trains of signals filtered in the range of five-minute periods, one can see that the line-width oscillations repeat the intensity signals in phase and amplitude with a high degree of accuracy (Figure~\ref{fig:wave-trains}).

From the analysis of the line-width and intensity oscillation characteristics, we can conclude that most likely both of these signals are a manifestation of the same MHD modes, since the periods and phases of the significant oscillations coincide in them.
This means that the observed oscillations in the line-width of the spectral line are not associated with the manifestations of torsional Alfv\'en waves.
The question remains open about the amplitude of the line-width oscillations, which is too large for the temperature oscillations caused by acoustic waves.

\citet{2015ApJ...799L..12D} provide another possible explanation for non-thermal variations of the line-profile widths.
This explanation suggests that non-thermal broadening of spectral lines can be caused by magnetoacoustic impacts propagating from below along vertical magnetic tubes.
In this case, the sawtooth shape of three-minute LOS-velocity oscillations in the chromosphere above a sunspot umbra indicates the presence of shock waves \citep{2006ApJ...640.1153C, 2014ApJ...786..137T, 2015LRSP...12....6K}, which, according to \citet{2015ApJ...799L..12D}, may cause periodic non-thermal broadening of spectral lines.
This suggestion needs a more detailed study of periodic three-minute variations of the spectral line-width variations above sunspot umbrae.

The waves that we observed in the lower atmosphere of coronal holes apparently cannot be classified as torsional Alfv\'en waves. We may assume that they are slow MHD waves.

\citet{2015NatCo...6.7813M}, observing the lower corona in the areas of open field lines in the \textit{Coronal Multi-Channel Polarimeter} (CoMP) data, noted an increased oscillation power in the three to five mHz range.
They attributed these oscillations to Alfv\'enic kink waves (propagating both upwards and downwards).
We observed signs of the wave propagation in a spectral range close to that of the oscillations observed in the lower corona in coronal holes, which are attributed to the manifestations of Alfv\'en waves.
\citet{2008SoPh..251..251C}, \citet{2012ApJ...751...31H}, and \citet{2015NatCo...6.7813M} showed mechanisms that may cause a transformation of the slow MHD waves ($p$-modes) into Alfv\'en waves.

Slow-mode waves are also found at coronal heights in coronal holes \citep{2011SSRv..158..267B,2014ApJ...789..118K}.
We assume that a part of the slow-mode waves from the lower atmosphere can undergo mode transformation in the upper chromosphere and serve as a source of Alfv\'en waves, which does not exclude the possibility of partial leaking without conversion.

\subsection{Estimating a Possible Input of Unresolved Flows in the Studied Signals}

The lower solar atmosphere -- especially, the chromosphere -- is a highly dynamic medium harboring, alongside with oscillations and waves, spontaneous plasma flows, which may demonstrate quasi-periodic behavior.
This makes them difficult to distinguish from purely wave processes in observations \citep{ 2010ApJ...722.1013D,2011ApJ...727...28T, 2012ApJ...759..144T}.
They are specifically challenging to differentiate in observations with a limited resolution. Such flows, however, may be identified in spectral observations by the asymmetry characteristics of the line profiles.
Recurrent variations in the line asymmetry resulting from quasi-periodic flows may add to the oscillations found in signals in LOS velocity and line-width signals.

To assess the impact of the non-wave dynamics on the signals that we use in the analysis, we studied the asymmetry of the line profiles in the observational series.
Fine unresolved flows may influence the shape of the lines, thus making an input in the LOS-velocity signals.
We measured the red--blue asymmetry profiles as in \citet{2011ApJ...738...18T}: we interpolated the line profiles ten times and subtracted the blue-wing intensity integrated over a narrow spectral range from that at the symmetrical position in the red wing roughly at the intensity level where the LOS velocity signals where taken.
Then we derived from these the variations that the changes in the asymmetry may have introduced to the velocity signals.
The measured asymmetry variations for the H$\alpha$ line are 1.5\,--\,2.2\,m{\AA}, which results in changes in the velocity signals of the order of 70\,--\,100\,m\,s$^{-1}$, while the typical amplitudes of the velocity signals in this line are 1100\,m\,s$^{-1}$.
A typical line-width oscillation amplitude in our analysis is 20\,--\,25\,m{\AA}.

This analysis suggests that the periodic changes in the asymmetry of the line profiles are much lower in magnitude than the line-width variations and LOS-velocity signals derived using the lambda-meter technique.
Thus, the unresolved flows in the aperture slit do not significantly influence these signals.

\section{Conclusion}

In this work, we analyzed the parameters of oscillations in the regions under coronal holes in the photosphere and chromosphere (the Si\,\textsc{i} 10827\,{\AA}\ and H$\alpha$ lines).

Compared to the quiet Sun, significant oscillations in a mutual range of periods was found in the coronal-hole regions at both studied levels. The range is a 1.6-minute wide band centered at 5.0 and 5.1 minutes for the first and second coronal holes.

Based on the phase-shift analysis, we observed predominantly upward propagation with an average phase shift of 22.6\,$\pm$\,12.8 degrees between the oscillations observed at the two levels.
This phase shift yields a propagation speed of 54\,km\,s$^{-1}$, which is close to the sound speed in the chromosphere.

The variations of the phase shift between the velocity and intensity signals in the lower atmosphere may indicate the presence of both standing and propagating waves over the time series. It is also possible that this variation is caused by the complications induced by non-wave phenomena such as spicules or jets.

In our data we tried to find manifestations of Alfv\'en waves under coronal holes.
As a proxy indicator of torsional Alfv\'en waves we used oscillations in the line width signals, whose frequencies found in our observations match those observed in the corona in open-field regions.
However, these line width oscillations seem to be associated with other MHD modes (we assume, slow MHD), since they accompany the intensity and LOS-velocity oscillations.
Nevertheless, physical mechanisms exist that allow both direct leakage and the transformation of the slow MHD waves that we observed in the lower atmosphere into Alfv\'en waves observed in the corona.

\acknowledgments

The reported study was funded by RFBR, project number 20-32-70076 and Project No.\,II.16.3.2 of ISTP SB RAS. Spectral  data were recorded at the Angara Multiaccess Center facilities at ISTP SB RAS. We acknowledge the NASA/SDO science teams for providing the data.
We thank the anonymous reviewer for helpful remarks.

\bibliography{Chelpanov}

\end{document}